# Title: Survival Function Analysis of Planet Orbit Semi-major Axis Distribution and Occurrence Rate Estimate

**Authors:** Li Zeng*[1], Stein B. Jacobsen[1], Dimitar D. Sasselov[2], Andrew Vanderburg[3]

**Affiliations:**

[1]Department of Earth & Planetary Sciences, Harvard University, 20 Oxford Street, Cambridge, MA 02138.

[2]Harvard-Smithsonian Center for Astrophysics, 60 Garden Street, Cambridge, MA 02138,

[3]Sagan Fellow, University of Texas at Austin

Correspondence to: astrozeng@gmail.com

**Abstract:**

Applying survival function analysis to the planet orbital period (**P**) and semi-major axis (**a**) distribution from the *Kepler* sample, we find that all exoplanets are uniformly distributed in (ln **a**) or (ln **P**), with an average inner cut-off of 0.05 AU to the host star. More specifically, this inner cut-off is 0.04 AU for rocky worlds (1-2 $R_\oplus$) and 0.08 AU for water worlds (2-4 $R_\oplus$). Moreover, the transitional planets (4-10 $R_\oplus$) and gas giants (>10 $R_\oplus$) have a change of slope of survival function at 0.4 AU from -1 to -1/2, suggesting a different statistical distribution uniform in $\sqrt{a}$ inside 0.4 AU, compared to small exoplanets (<4 $R_\oplus$). This difference in distribution is likely caused by the difference in planet migration mechanism, and susceptibility to host stellar irradiation, for gas-poor (<4 $R_\oplus$) versus gas-rich (>4 $R_\oplus$) planets. Armed with this knowledge and combined with the survival function analysis of planet size distribution, we can make precise estimates of planet occurrence rate and predict the *TESS* mission yield.

**Method:**

**(1) Survival Function Analysis**

The survival function (Clauset, Rohilla Shalizi, & J Newman, 2009; Feigelson & Nelson, 1985; Virkar & Clauset, 2014), also known as the complimentary cumulative distribution function (cCDF), is defined as the number of planets above a given parameter (here Period P or semi-major axis a) versus that parameter, in a log-log plot, of the *Kepler* planet candidates: 4433 from Q1-Q17 *NASA* Exoplanet Archive (Akeson et al., 2013), with false positives excluded already.

SF (Survival Function) = 1- CDF (Cumulative Distribution Function) = 1 - Integral of PDF (Probability Density Function)
Differentiate this SF, one gets the PDF (Probability Density Function).

The survival function (SF) can tell apart different distributions. Comparing to the probability density function (PDF), it has the advantage of overcoming the large fluctuations that occur in the tail of a distribution due to finite sample sizes (Clauset et al., 2009). For example, on a log-log plot of survival function, power-law distribution appears as a straight line, while normal, log-normal, or exponential distributions all have a sharp cut-off (upper bound) in x-axis. This plot is also known as the rank-frequency plot (Newman, 2005). The comparison of the SF of the data with the straight-line SF of reference power-law distributions in **Figure 1-3** are the essence of the Kolmogorov-Smirnov (K-S) test. K-S statistic simply evaluates the maximum distance between the CDFs of the data and the fitted model and can also test the goodness-of-fit. This approach identifies the boundaries separating different regimes of probabilistic distributions in the data.

**Discussion:**

**Part I. Survival function analysis of the overall planet population**

**(1) survival function of orbital period**

**Figure 1** shows the SF of the orbital period (**P**) of all 4433 *Kepler* planet candidates. The bulk of it is best fit to a power-law, with power index equal to **-2/3**: SF ∝ **P^(-2/3)**. Differentiating this SF gives the probability density function: PDF∝ **P^(-5/3)**. Thus, the likelihood **dN** of finding a planet around one host star in the interval of (P~P+dP) is **dN ∝ P^(-5/3) dP**.

So far, this PDF does not represent the true distribution of planets, until we make completeness corrections for **(1) geometric transit probability** and **(2) pipeline incompleteness**.

**(2) geometric transit probability**

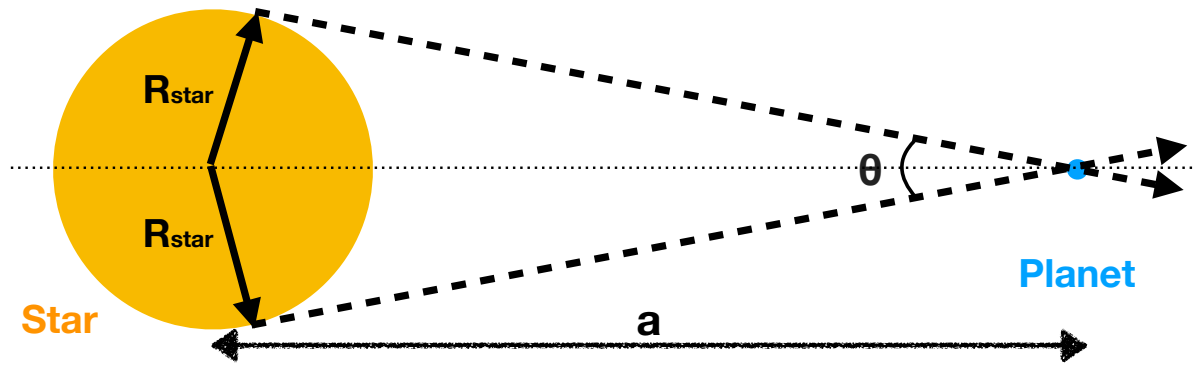

The **geometric correction** has to do with that the more distant planets have less chance of being aligned with our line-of-sight and thus less chance of transiting, see illustration above. If we assume the orientation of orbital plane in space is random, then this **probability of transit** = $\frac{(R_{star}+R_{planet})}{a}$, according to (Winn, 2010). $R_{planet} \ll R_{star}$, and $R_{star} \approx R_{sun}$, because most host stars in the *Kepler* catalog are solar-type FGK main-sequence stars. So the **probability of transit** ≈ $\frac{R_{sun}}{a}$ ∝ **a^(-1)** ∝ **P^(-2/3)**. To correct for this **probability of transit** to back out the Survival Function Analysis of Planet Orbit Distribution: Zeng, Jacobsen, Sasselov, Vanderburgtrue distribution, the PDF needs to be multiplied by a factor of $\frac{a}{R_{sun}} \propto a \propto P^{2/3}$. Here we refer to the transit-probability-corrected distribution as **$dN_{corrected} \propto P^{-1} dP \propto d \log P$**.

**(3) pipeline incompleteness**

The pipeline incompleteness causes incomplete detection of planets of radius less than a certain radius threshold. This radius threshold depends on the orbital period considered. The shorter the orbital period, the more transits are observed within a given time frame and the higher the signal-to-noise ratio as a result of that. Given the same instrument sensitivity, this period-dependent threshold in planet radius is estimated for Kepler pipeline, above which the detection is mostly complete, and below which the detection is very incomplete. It varies between 1~2 $R_\oplus$ for orbital period range of 1~300 day, and for most Kepler planets within 30-day orbit, this threshold is ~1.2 $R_\oplus$, see Figure 6 of (Fulton et al., 2017). The exact number also varies depending on which version of pipeline one is using. See also Figure 2 of KSCI-19111-001 (Christiansen & L., 2017) and Figure 7 of KSCI-19111-002 (Burke & Catanzarite, 2017).

**(4) survival function of semi-major axis**

**Figure 2** shows the SF of the orbital semi-major axis (**a**) for all 4433 *Kepler* planet candidates. SF $\propto a^{-1}$, PDF $\propto a^{-2}$, $dN \propto a^{-2} da$, and $dN_{corrected} \propto a^{-1} da \propto d \log a$. This result is consistent with the result of **Figure 1** since $a^3 \propto P^2$ (the *Kepler*'s third law). In short, we can re-write $dN_{corrected} \propto d \log a \propto d \log P$.

Here, we can identify the **inner cut-off** of the power-law relation in both **Figure 1** and **Figure 2** by pinpointing the intersection of the fitted power-law straight line with that of the horizontal saturation line in the log-log rank-frequency plot. This inner cut-off is **P=4-day** orbital period for all *Kepler* planet candidates on average, as shown in **Figure 1**. Applying the *Kepler*'s third law for 1 solar mass host-star, the corresponding semi-major axis **a** = $(4/365.25)^{2/3}$ = **0.05 AU**. It matches exactly the inner cut-off identified in **Figure 2**. This agreement acts as a check for the validity of this method.

This inner cut-off of 0.05 AU means that most planets cannot get closer than this distance to the host star due to physical limitations. Knowing that 1 AU = 215 solar radii, then 0.05 AU ≈ 10 solar radii. Planets inside 10 solar radii or 10 stellar radii are highly irradiated, and experience strong tidal interactions with their host stars. All these can lead to short lifetimes for these planets if they get that close.

There is an **outer cut-off** of the power-law relation in **Figure 1** and **Figure 2** due to the limited observational duration of the *Kepler* mission. The *Kepler* telescope was operational since December 2009 until the failure of its second reaction wheel in May 2013. The entire duration of its normal operation is 3.5 year for the data concerned in this paper, not including K2. Since it generally requires three transits to confirm one planet candidate. Thus, beyond ~3.5/3.5=1 year or ~1 AU orbit, *Kepler*'s detection becomes very incomplete, as most detections have less than three repeats and cannot be listed as candidates. This can be seen as a rapid drop-off of SF if plotted beyond 1 AU.

Extrapolating the true distribution beyond this **outer cut-off**, planets are expected to be uniformly-distributed in **Log a** or **Log P** towards larger orbital distances. This expectation is in accordance with the *Titius-Bode Law*: a modified version of a power-law uniform in **Log a** or **Log P**. It can explain all the orbital distances of major planetary objects in our own solar system. From this point of view, although some *Kepler* systems are one-order-of-magnitude more compact than our own solar system, their relative planet spacing is similar. This seems to suggest uniform in **Log a** or **Log P** is the natural consequence or the primordial configuration of planet formation, and the compactness is a result of different degrees of sub-sequent migrations.

**(5) statistics of planet occurrence rate**

There are 145,000 main-sequence stars, most of them sun-like FGK stars, being continuously monitored by the *Kepler* mission over its 3.5-year duration of normal operation. Out of them about 4000 planet candidates with radii larger than 1 Earth radius (>1 $R_\oplus$) have been identified. With this information, we can obtain the occurrence rate of planets (>1 $R_\oplus$) around sun-like stars.

Previously, we derive: $dN_{corrected} = A \times d(\ln a)$, where $dN_{corrected}$ represents the transit-probability-corrected average number of planets per host star within a given range of semi-major axis (**a~a+da**). Here A is the dimensionless normalization factor. Its integral gives:

$N_{corrected} (a_{in} \sim a_{out}) = \int_{a_{in}}^{a_{out}} dN_{corrected} = A \times \ln\left(\frac{a_{out}}{a_{in}}\right)$

This formula is useful in estimating the number of planets in the habitable zone: once the inner and outer bound of habitable zone are given, then $N_{corrected} (a_{in} \sim a_{out})$ can be directly calculated.

Similarly, the average observed number of planets per host star within (**a~a+da**) is:
$dN_{obs} \approx A \times \frac{R_\odot}{a} \times d(\ln a) = A \times \frac{R_\odot}{a} \times \frac{da}{a} = A \times R_\odot \times \frac{da}{a^2}$

$N_{obs} (a_{in} \sim a_{out}) = \int_{a_{in}}^{a_{out}} dN_{obs} = \int_{a_{in}}^{a_{out}} A \times \frac{R_\odot}{a} \times \frac{da}{a} = A \times R_\odot \times \left(\frac{1}{a_{in}} - \frac{1}{a_{out}}\right)$

Plug in the the inner cut-off ($a_{in}=0.05$ AU) and the outer cut-off ($a_{out}=1$ AU), the average number of planets ($N_{obs}$) observed per host star is estimated as:

$N_{obs} (0.05\text{AU} \sim 1\text{AU}) = A \times R_\odot \times \left(\frac{1}{a_{in}} - \frac{1}{a_{out}}\right)\Big|_{0.05\text{AU}}^{1\text{AU}} \approx A \times R_\odot \times \frac{1}{a_{in}}$

Equating this to the number of planets observed, we can find A as:

$\frac{4,000}{145,000} = N_{obs} = \frac{A \cdot R_\odot}{a_{in}} = A \cdot \frac{1}{215 \cdot 0.05} \Rightarrow A_{(>1R_\oplus)} = \frac{215}{\left(\frac{1}{0.05}\right)} \cdot \frac{4,000}{145,000} = 0.3$

The result suggests that, beyond the inner cut-off of 0.05 AU, one expects to find on average about 0.3 planet with radii >1 $R_\oplus$, per unit natural-logarithmic bin of a, per star, up to at least 1 AU. This population of planets is dominated by small planets in between 1 $R_\oplus$ and 4 $R_\oplus$.

**Part II. Survival function analysis of separate planet groups**

**(1) Planets in different radius bins have different SF slope and inner cut-off.**

This analysis can be applied to planets divided into four radius bins (1-2 $R_\oplus$, 2-4 $R_\oplus$, 4-10 $R_\oplus$, and >10 $R_\oplus$), where the bins are identified through a model-independent survival function analysis of planet radii in a separate paper (Zeng, Jacobsen, Sasselov, & Vanderburg, 2018). The results are shown in **Figure 3**.

The most prominent feature is that small (gas-poor) exoplanets (<4 $R_\oplus$) have a different power-law dependence on orbital semi-major axis (a) than the large (gas-rich) exoplanets (>4 $R_\oplus$) within 0.4 AU:
For small exoplanets (1-4 $R_\oplus$), $dN_{corrected} \propto a^{-1} da$
For large exoplanets (>4 $R_\oplus$), $dN_{corrected} \propto a^{-1/2} da$

This difference in power-law dependence inside 0.4 AU can result from different planet-migration mechanism and/or susceptibility to host stars' influence, for small gas-poor (<4 $R_\oplus$) planets versus large gas-rich (>4 $R_\oplus$) planets. Gas giants (>10 $R_\oplus$) have a break in their SF slope at 0.4 AU, where the SF slope changes from -1 to -1/2. This suggests there is a relative depletion of them inside 0.4 AU. However, this cut-off is incomplete, so that a fraction of them can approach their host stars to extreme proximity (0.02 AU or about 4 solar or stellar radii from the center of the star).

The transitional planets (4-10 $R_\oplus$) have SF somewhat similar to that of gas giants (>10 $R_\oplus$) but smoother. They start to deviate from the small planets' -1 slope at around 0.3~0.4 AU and have absolute inner cut-off at around 0.05 AU.

Statistically speaking, the inner cut-off of planets of (2-4 $R_\oplus$) is 0.08 AU, which is twice that of the inner cut-off of planets of (1-2 $R_\oplus$) at 0.04 AU. Otherwise, they have the same power-law dependence beyond their inner cut-offs. Correcting for the geometric probability of transit, even though there appears to be about equal number of planet candidates observed for each of the two bins, the number of planets of (2-4 $R_\oplus$) is about twice the number of planets of (1-2 $R_\oplus$) since they are twice the distance away from their host star on average and the transit probability is half. Pipeline incompleteness correction could bring their numbers closer.

**(2) Occurrence rate of planets in each radius bin: (1-2 $R_\oplus$), (2-4 $R_\oplus$), (4-10 $R_\oplus$), (>10 $R_\oplus$)**

**(1-2 $R_\oplus$) rocky worlds**: inner cut-off $a_{in}$=0.04 AU, $dN_{corrected} = A_{(1-2\ R_\oplus)} \times d(\ln a)$:

$$A_{(1-2\ R_\oplus)} \approx \frac{215}{\left(\frac{1}{0.04}\right)} \cdot \frac{1{,}728}{145{,}000} = 0.1$$

That is, on average, we expect to find about 0.1 planet of (1-2 $R_\oplus$) per unit natural-logarithmic interval of semi-major axis (a) outside the inner cut-off of 0.04 AU.

**(2-4 $R_\oplus$) water worlds**: inner cut-off $a_{in}$=0.08 AU, $dN_{corrected} = A_{(2-4\ R_\oplus)} \times d(\ln a)$:

$$A_{(2-4\,R_\oplus)} \approx \frac{215}{\left(\frac{1}{0.08}\right)} \cdot \frac{1{,}559}{145{,}000} = 0.2$$

Thus, we expect to find about 0.2 planet of (2-4 $R_\oplus$) per unit natural-logarithmic interval of semi-major axis (a) outside the inner cut-off of 0.08 AU.

For planets of (4-10 $R_\oplus$) and (>10 $R_\oplus$), a different normalization is needed since their SF slope is close to -1/2 inside 0.4 AU. Introduce the normalization factor B: (where B has the dimension of $1/\sqrt{AU}$)

$$dN_{\text{corrected}} = B \times d(\sqrt{a})$$

For a given interval of semi-major axis $a_{in}$ and $a_{out}$,

$$N_{\text{obs}} = \int_{a_{in}}^{a_{out}} dN_{obs} = \int_{a_{in}}^{a_{out}} \frac{R_\odot}{a} \cdot dN_{corrected} = \int_{a_{in}}^{a_{out}} \frac{R_\odot}{a} \times B \times d\sqrt{a} = B \times R_\odot \times \left(\frac{1}{\sqrt{a_{in}}} - \frac{1}{\sqrt{a_{out}}}\right)$$

**(4-10 $R_\oplus$) transitional planets**: $a_{in}$=0.05 AU, $a_{out}$=0.3 AU,

$$B_{(4-10\,R_\oplus)} \approx \frac{215}{\left(\frac{1}{\sqrt{0.05}} - \frac{1}{\sqrt{0.3}}\right)} \cdot \frac{358 - 150}{145{,}000} = 0.1 * \frac{1}{\sqrt{AU}}$$

**(>10 $R_\oplus$) gas giants**: $a_{in}$=0.02 AU, $a_{out}$=0.4 AU,

$$B_{(>10\,R_\oplus)} \approx \frac{215}{\left(\frac{1}{\sqrt{0.02}} - \frac{1}{\sqrt{0.4}}\right)} \cdot \frac{300 - 77}{145{,}000} = 0.06 * \frac{1}{\sqrt{AU}}$$

And,

$$N_{\text{corrected}} (<a_{out}) = B \times (\sqrt{a_{out}} - \sqrt{a_{in}})$$

Thus,

$$N_{\text{corrected},(4-10\,R_\oplus)}(< 0.4\text{AU}) \approx B_{(4-10\,R_\oplus)} \times \left(\sqrt{0.4\text{AU}} - \sqrt{0.05\text{AU}}\right) = 0.04$$
$$N_{\text{corrected},(>10\,R_\oplus)}(< 0.4\text{AU}) \approx B_{(>10\,R_\oplus)} \times \left(\sqrt{0.4\text{AU}} - \sqrt{0.02\text{AU}}\right) = 0.03$$

In total, there is about (0.04+0.03)=0.07 planet of (>4 $R_\oplus$) per host star within 0.4 AU. In comparison, there is about 0.6 planet of (<4 $R_\oplus$) within 0.4 AU around a typical FGK-star. Thus, large planets (>4 $R_\oplus$) are depleted inside 0.4 AU compared to small planets (<4 $R_\oplus$).

Beyond 0.4 AU, both planets of (4-10 $R_\oplus$) and (>10 $R_\oplus$) show similar power-law dependence on semi-major axis (a) as planets of (1-2 $R_\oplus$) and (2-4 $R_\oplus$). So **dN$_{\text{corrected}}$** ∝ d (ln **a**) ∝ d (ln **P**) is likely the primordial/generic configuration of a planet system from formation, regardless of planet sizes. Large planets (>4 $R_\oplus$) are depleted inside 0.4 AU. Beyond 0.4 AU, the normalization factor A for planets of (4-10 $R_⊕$) and (>10 $R_⊕$) where $dN_{corrected} ≈ A × d(\ln a)$:

$$A_{(4-10\ R_⊕)} ≈ \frac{215}{\left(\frac{1}{0.3}\right)} · \frac{150}{145,000} = 0.067$$

$$A_{(>10\ R_⊕)} ≈ \frac{215}{\left(\frac{1}{0.4}\right)} · \frac{77}{145,000} = 0.045$$

In summary, the intrinsic relative abundance ratio of the four planet types is:

$$A_{(1-2\ R_⊕)} : A_{(2-4\ R_⊕)} : A_{(4-10\ R_⊕)} : A_{(>10\ R_⊕)} = (0.1) : 0.2 : 0.067 : 0.045$$

Where (0.1) is in parentheses because it is an under-estimate subject to pipeline incompleteness correction of smallest exoplanets.

$$A_{(1-2\ R_⊕)} : A_{(2-4\ R_⊕)} : A_{(4-10\ R_⊕)} : A_{(>10\ R_⊕)} ≈ \left(\frac{1}{2}\right) : 1 : \frac{1}{3} : \frac{1}{4}$$

**Conclusion**

Planets are intrinsically uniformly-distributed in $(\ln a)$ or $(\ln P)$, as **$dN_{corrected} = A × d(\ln a)$**, with a dimensionless normalization constant A and an inner cut-off $a_{in}$.

For the overall planet population of (>1 $R_⊕$), $A_{(≳1 R_⊕)} = 0.3$ and $a_{in} = 0.05$ AU.
If dividing planets into different radius bins as (Zeng et al., 2018):

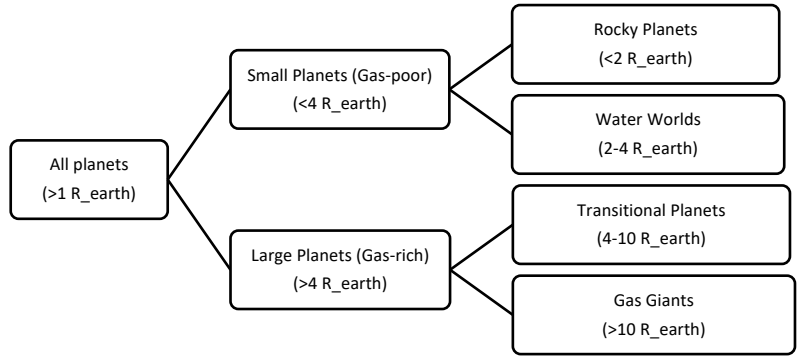

Then,
**(1-2 $R_⊕$) rocky planets**: $A_{(1-2\ R⊕)} = 0.1$, $a_{in} = 0.04$ AU
**(2-4 $R_⊕$) water worlds**: $A_{(2-4\ R⊕)} = 0.2$, $a_{in} = 0.08$ AU
**(4-10 $R_⊕$) transitional planets**: $A_{(4-10\ R⊕)} = 0.067$, $a_{in} = 0.4$ AU
**(>10 $R_⊕$) gas giants**: $A_{(>10\ R⊕)} = 0.045$, $a_{in} = 0.4$ AU

Inside 0.4 AU, large planets (>4 $R_⊕$) shows a different distribution: **$dN_{corrected} ≈ B × d(\sqrt{a})$**
**(4-10 $R_⊕$) transitional planets**: $B_{(4-10\ R⊕)} ≈ 0.04/\sqrt{AU}$, $a_{in} = 0.05$ AU

**(>10 $R_⊕$) gas giants**: $B_{(>10\ R⊕)} \approx 0.03/\sqrt{AU}$, $a_{in}$=0.02 AU

In general, large planets (>4 $R_⊕$) are depleted inside 0.4 AU. On average, there is about 0.07 planet of (>4 $R_⊕$) within 0.4 AU around a typical FGK-star. Comparatively, there is about 0.6 planets of (<4 $R_⊕$) within 0.4 AU around a typical FGK-star.

Small exoplanets (<4 $R_⊕$) have inner cut-offs of closest distances to their host stars. Planets of (2-4 $R_⊕$) has inner cut-off at 0.08 AU (20 solar radii away, or about 1000 K surface equilibrium temperature with Earth-like albedo), about twice the distance as the inner cut-off of planets of (1-2 $R_⊕$) at 0.04 AU (10 solar radii away, or about 1400 K surface equilibrium temperature with Earth-like albedo).

Gas giants (>10 $R_⊕$) exhibit two cut-offs or changes of slope. The first one is at 0.4 AU, where the survival function slope changes from -1 to -1/2. The second one is at 0.02 AU or about 2 Roche radii away from the host star.

The simplest explanation for slopes in **Figure 3** is that, -1 slope is a result of migration due to planet-disk interaction, and -1/2 slope is a result of migration due to planet-planet scattering (Nelson, Ford, & Rasio, 2017).

**Acknowledgements**: This work was partly supported by a grant from the Simons Foundation (SCOL [award #337090] to L.Z.). Part of this research was also conducted under the Sandia Z Fundamental Science Program and supported by the Department of Energy National Nuclear Security Administration under Award Numbers DE-NA0001804 and DE-NA0002937 to S. B. Jacobsen (PI) with Harvard University. This research is the authors' views and not those of the DOE. Sandia National Laboratories is a multimission laboratory managed and operated by National Technology and Engineering Solutions of Sandia, LLC., a wholly owned subsidiary of Honeywell International, Inc., for the U.S. Department of Energy's National Nuclear Security Administration under contract DE-NA-0003525.

Survival Function Analysis of Planet Orbit Distribution: Zeng, Jacobsen, Sasselov, Vanderburg

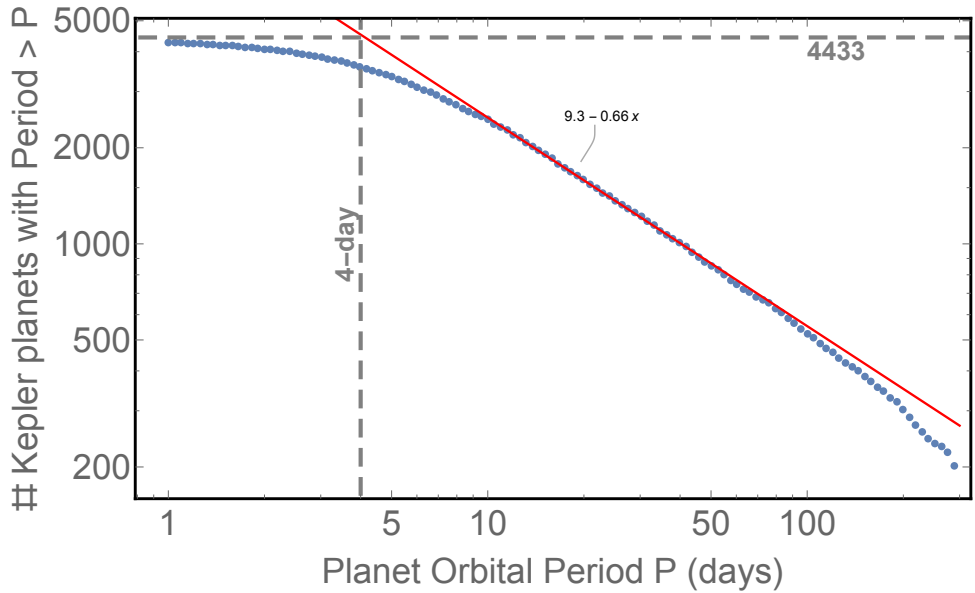

**Figure 1.** Survival function of planet orbital period P (days) for *Kepler* planet candidates (4433 from Q1-Q17 *NASA* Exoplanet Archive (Akeson et al., 2013), false positives excluded already). The best fit to power-law is in natural-logarithm shown as the red line. The slope of this power-law is exactly 2/3 in log-log plot. The inner cut-off is 4-day orbital period, as identified by the intersection of the extrapolation of the power-law with the horizontal line representing the total number of planets in this case.

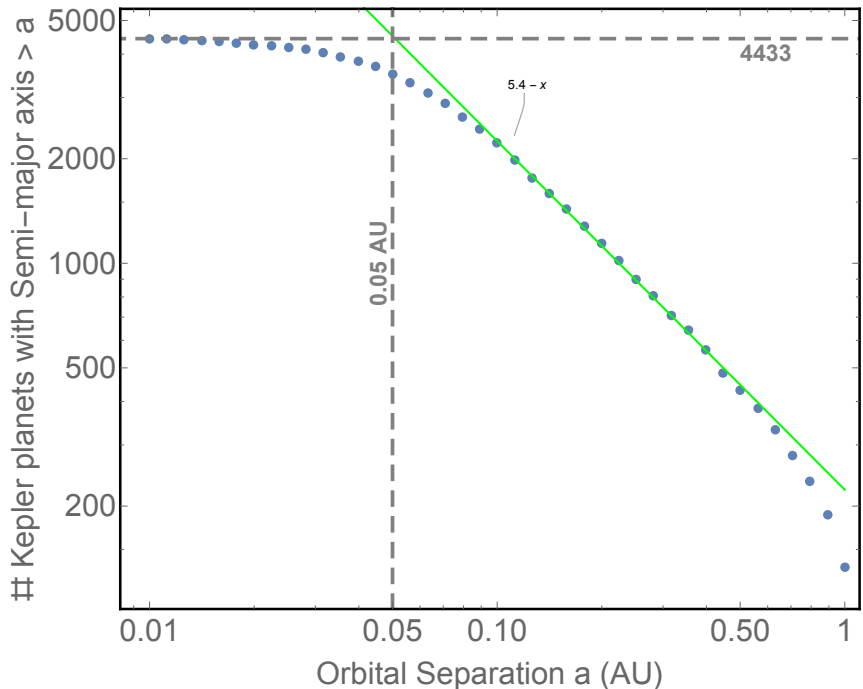

**Figure 2.** Survival function of planet semi-major axis a (AU) for *Kepler* planet candidates (4433 from Q1-Q17 *NASA* Exoplanet Archive (Akeson et al., 2013), false positives excluded already). The best fit to power-law is in natural-logarithm shown as the green line. The inner cut-off is identified by the intersection of the extrapolation of the power-law with the horizontal line representing the total number of planets in this case. It is 0.05 AU, corresponding to exactly 4-day orbital period for a planet around 1 solar mass star, as expected, because most host stars in *Kepler* sample are solar type FGK stars with masses close to 1 solar mass.

Survival Function Analysis of Planet Orbit Distribution: Zeng, Jacobsen, Sasselov, Vanderburg

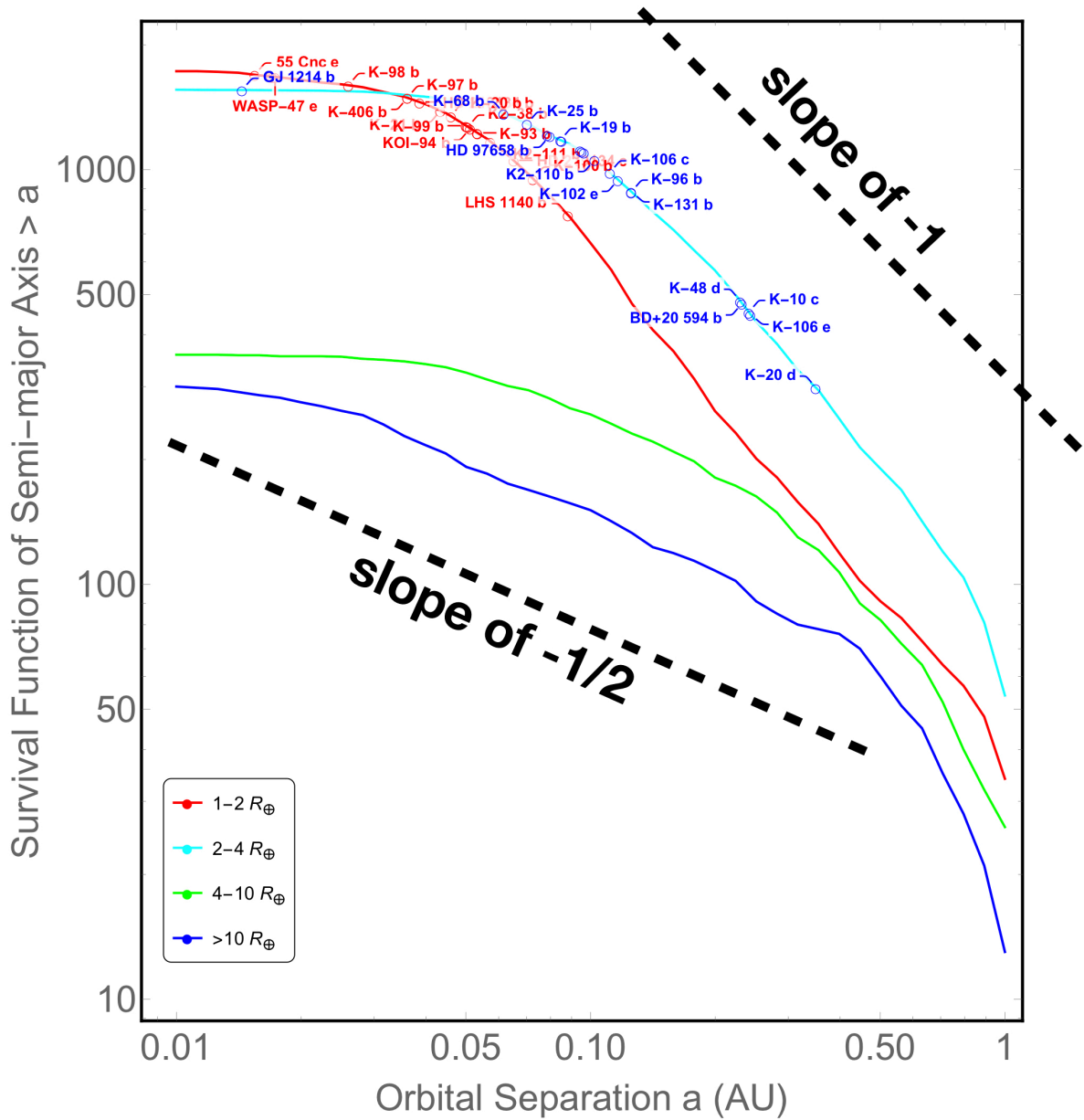

**Figure 3.** Survival function analysis of *Kepler* planet candidates (4433 from the Q1-Q17 *NASA* Exoplanet Archive (Akeson et al., 2013), false positives excluded already), divided in four radius bins: red: 1-2 $R_\oplus$, cyan: 2-4 $R_\oplus$, green: 4-10 $R_\oplus$, and blue: >10 $R_\oplus$. This classification scheme based on planet radii is adopted from the survival function analysis of planet radius distribution (Zeng et al., 2018). The straight lines are power-law fits expressed in natural logarithm.